\documentclass[nohyperref]{article}

\usepackage{microtype}
\usepackage{graphicx}
\usepackage{subfigure}
\usepackage{booktabs} 

\usepackage{hyperref}

\usepackage[accepted]{icml2022}

\usepackage{amsmath}
\usepackage{amssymb}
\usepackage{amsbsy}
\usepackage{mathtools}
\usepackage{amsthm}
\usepackage{paralist}
\usepackage{paralist}
\usepackage[capitalize,noabbrev]{cleveref}
\usepackage{stix,bbding,pifont,utfsym,fontawesome}
\theoremstyle{plain}

\theoremstyle{definition}

\theoremstyle{remark}

\usepackage[textsize=tiny]{todonotes}

\icmltitlerunning{Comparing supervised and self-supervised embedding for ExVo Multi-Task learning track }

\begin{document}

\twocolumn[
\icmltitle{Comparing supervised and self-supervised embedding for ExVo Multi-Task learning track }



\icmlsetsymbol{equal}{*}

\begin{icmlauthorlist}
\icmlauthor{Tilak Purohit}{comp,sch}
\icmlauthor{Imen Ben Mahmoud}{comp}
\icmlauthor{Bogdan Vlasenko}{comp}
\icmlauthor{Mathew Magimai.-Doss}{comp}
\end{icmlauthorlist}

\icmlaffiliation{comp}{Idiap Research Institute, Martigny, Switzerland}
\icmlaffiliation{sch}{Ecole polytechnique f\'ed\'erale de Lausanne (EPFL), Switzerland}

\icmlcorrespondingauthor{Tilak Purohit}{tilak.purohit@idiap.ch}
\icmlcorrespondingauthor{Mathew Magimai.-Doss}{mathew@idiap.ch}

\icmlkeywords{Machine Learning, ICML}

\vskip 0.3in
]



\printAffiliationsAndNotice{}  

\begin{abstract}
The ICML Expressive Vocalizations (ExVo) Multi-task challenge 2022, focuses on understanding the emotional facets of the non-linguistic vocalizations (vocal bursts (VB)). The objective of this challenge is to predict emotional intensities for VB, being a multi-task challenge it also requires to predict speakers' age and native-country. For this challenge we study and compare two distinct embedding spaces namely, self-supervised learning (SSL) based embeddings and task-specific supervised learning based embeddings. Towards that, we investigate feature representations obtained from several pre-trained SSL neural networks and task-specific supervised classification neural networks.
Our studies show that the best performance is obtained with a hybrid approach, where predictions derived via both SSL and task-specific supervised learning are used. Our best system on test-set surpasses the ComPARE baseline (harmonic mean of all sub-task scores i.e., $S_{MTL}$) by a relative 13\% margin.
\end{abstract}

\vspace*{-1cm}
\section{Introduction}
\label{sec:intro}
\vspace*{-0.2cm}
It is well known that human speech is a rich source of emotional signals~\cite{darwin2015expression}, apart from verbal-speech humans also express emotions through non-verbal expressive vocalizations (ExVo) such as laughter, cries and gasp~\cite{bachorowski2001acoustic,simon2009voice}. In the past two decades there has been a tremendous surge of interest in speech emotion recognition (SER) research~\cite{schuller2018speech}, for there are several applications of SER in the field of affective computing to human-computer interfaces. Typically, speech emotion studies have been conducted on human verbal speech, recorded from dyadic communications to acted or scripted scenarios~ \cite{burkhardt2005database, busso2008iemocap, engberg1996documentation}. 
However, literature on modelling/quantifying the non-verbal ExVo aspects for the emotion recognition (ER) is sparse. Modeling ExVo could help in filling the research gaps in speech emotion research and may help in improving the current SER systems, hence, understanding these non-verbal ExVo becomes important. ICML ExVo 2022 challenge~\cite{BairdExVo2022} for the first time provides the \textsc{HUME-VB} corpus~\cite{Cowen2022HumeVB} which allows the modelling of granular emotion characteristics of vocal bursts (VB). This corpus enables to make a distinction between the emotion of a VB, for example the VB for laughter could have emotion like amusement, awkwardness, and triumph associated to it. Similarly, VB for gasps could elicit emotions like awe, excitement, fear, and surprise, and VB for cries may produce emotions like distress, horror, and sadness. 

\begin{figure}[!htb]
	\centering
	\includegraphics[trim={0cm 0cm 0cm 0cm},clip=true,width=0.4\textwidth]{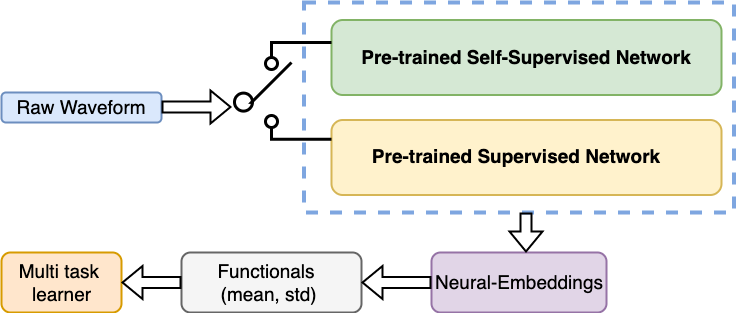}
	\vspace*{-0.3cm}
	\caption{Proposed neural embedding based approaches for ExVo Multi-task learning.}
	\label{fig:sup_approach_combined}
	\vspace*{-0.3cm}
\end{figure}

In this paper, we focus on the first track of the challenge, ExVo Multi-Task learning. Taking inspirations from recent works on the use of embeddings from the pre-trained networks for various speech procession tasks including paralinguistic tasks  \cite{keesing2021acoustic,yang2021superb,mostaani2022modeling,srinivasan2022representation}, we investigate the utility of neural embeddings for speakers' emotion intensity, native country and age estimation. In that regard, as illustrated in Figure~\ref{fig:sup_approach_combined}, we compare two types of neural embedding extraction approaches: (a) neural embeddings extracted from neural networks trained in self-supervised learning (SSL) setting and (b) neural embeddings extracted from neural networks trained on auxiliary out-of-domain tasks such as, SER, phone classification and on in-domain ExVo challenge task.
\vspace*{-0.3cm}

The paper is structured as follows. 
Section~\ref{sec:Experimental_Setup} describes our experimental setup, derived features and multi-task learning framework. We present our results and the analysis in Section~\ref{sec:Results}. Section~\ref{sec:conclusion} concludes the paper.

\vspace*{-0.3cm}
\section{Experimental Setup}
\label{sec:Experimental_Setup}
\vspace*{-0.2cm}
In this section we present the ExVo 2022 challenge dataset and the experiment protocols. Next we present the fixed-length feature representation obtained through the approaches we defined in Section~\ref{sec:intro}, and finally the multi-task learner block trained for the task at hand.
\vspace*{-0.3cm}
\subsection{Database and Protocols}
\vspace*{-0.2cm}
The ExVo competition dataset \cite{Cowen2022HumeVB} contains 59,201 recordings of VB from 1,702 speakers, with approximately 37 hours of 16-bit audio sampled at 16kHz (we used single channel for our study). The recordings were collected from four different countries, namely-  USA, China, Venezuela, and South Africa from the speakers with age ranging between 18 to 39 years old. The data was collected at speakers' own home from a microphone hence the data is considered `in-the-wild'. Each VB has intensity scores (ranging from $[0,100]$) corresponding to 10 different classes of emotion: amusement, awe, awkwardness, distress, excitement, fear, horror, sadness, surprise and triumph. These scores were averaged across 85 raters and further were normalised to a range from $[0,1]$. 
For the ExVo Multi-Task challenge, the data was split equally into 3 sets - train, validation and test, where labels were provided corresponding to train and validation set only. Each team had five trials to evaluate the developed systems.
The predictions were evaluated based on task-specific metrics: 
\begin{compactenum}
    \item \textit{Concordance Correlation Coefficient} (CCC): for the emotion recognition task. Calculated for every emotion, the overall \textit{CCC} is defined as the mean of the coefficients.
    \item The \textit{Mean Absolute Error} (MAE): is calculated for the age detection.
    \item  \textit{The Unweighted Average Recall} (UAR): is reported for the native country classification task.
\end{compactenum}
The overall prediction score is the \textit{Harmonic mean}, defined as follows,
$S_{MTL}= \frac{3}{(1/CCC+ MAE + 1/UAR)}$  \nonumber 
. We followed the protocols as set by the organizers to train and evaluate our systems.
\subsection{Extracting neural embedding-based fixed-length feature representations}

For deriving SSL embeddings, we make use of the following publicly available state-of-the-art (SOTA) pre-trained SSL systems like: Wav2Vec2 \cite{baevski2020wav2vec},  HuBERT \cite{hsu2021hubert_emb}, and WavLM \cite{WavLM}. These systems are among the top three performing networks for the SUPERB challenge \cite{yang2021superb}, a SSL benchmark challenge for the speech processing tasks.       


The task-specific embeddings were derived using pre-trained supervised networks. To enrich the study we took several systems, three out-of-the-domain trained systems and four systems trained on ExVo data that is in-domain trained network. The out-of-domain task-specific trained systems and the motivation to consider them are as following:     

\underline{(1) Raw (SER)}: We used an off-the-shelf CNN network that was trained on IEMOCAP \cite{busso2008iemocap} dataset to classify four emotion categories namely anger, happy, sad, neutral in an end-to-end fashion by taking 250ms of raw-speech signal as input. This system is chosen since its trained for SER task on sub-segmental speech (typically 250ms) and may help with ExVo ER sub-task. The network comprises of four convolution layers followed by one hidden layer with ten nodes and an output layer corresponding to four emotion classes with a softmax activation function, while the other layers are followed by a ReLU activation. The network was trained in a speaker-independent manner and fetched an unweighted average recall (UAR) of 57.4\%. Using this network frame-level neural embeddings of dimension 10, are extracted before the activation of last hidden layer. A fixed-length utterance-level feature representation is obtained by computing functionals (mean and standard deviation) of the frame-level neural embeddings, denoted by Raw(SER). This makes the Raw(SER) embedding of dimension 20 (10 mean $+$ 10 std).  \\
\underline{(2) Zff (SER)}: The system is similar to the above mentioned network in all the aspects (like input signal duration, training dataset, architecture, etc.) but instead of raw-speech waveform the network is trained on speech signal after filtering it through a zero frequency filter (Zff) \cite{murty2009characterization} which models the speech-source related information, and here the input signal is zff-waveform. This model was chosen since it has been shown to improve the performance for paralinguistic tasks like, depression detection \cite{dubaguntaicassp2019} and dementia detection \cite{cummins2020comparison} when compared to only speech signal. The model achieved an UAR of 48.4\% on a four class emotion classification task on IEMOCAP dataset. Similar to the above system we derive utterance-level embeddings denoted by Zff(SER) with dimension 20. \\
\underline{(3) Raw (ASR)}: We took an off-the-shelf CNN based network that models raw-waveform for phone classification. The network consists of 10 convolutional layers followed by a fully connected layer with 1024 nodes and an output layer. The network was originally trained on AMI corpus \cite{carletta2007unleashing}. The frame level neural embeddings of dimension 1024 are converted to fixed-length utterance-level embeddings using functionals (mean and std), denoted by Raw(ASR). The final embedding dimension is 2048 (1024 for mean + 1024 for std). Since this system was trained on a conversational dataset, the embeddings might capture country specific articulation which may help with the ExVo country recognition sub-task. Also, emotion information has been found to be carried at phonetic level~\cite{Vlasenko11_vowels,yuan2021role}.        

\begin{table}[!t]
    \centering 
    \footnotesize{
    \begin{tabular}{ccccc} 
{\bf Systems} & {\bf Classifier} & {\bf UAR} & {\bf WAR} \\ \hline
Raw(ExVo)-ER & Softmax & 28.51 & 43.97 \\
Zff(ExVo)-ER & Softmax & 20.79 & 34.32\\ 
\hline
Raw(ExVo)-CR & Softmax & 43.70 & 50.01 \\
Zff(ExVo)-CR & Softmax & 45.65 & 51.46 \\ 
\hline
\hline
\end{tabular}
}
\caption{Classification based performance on ExVo validation-set in terms of UAR and Weighted average recall (WAR). Abbreviations: ER - emotion recognition, CR - country recognition}
\vspace*{-0.7cm}
\label{tab:classification_results}
\end{table}    

\underline{(4) Raw/Zff(ExVo)-ER}: 
We developed an end-to-end system which takes 250ms of raw (/zff)-waveform as input. The system was trained for a ten-class classification task using the ExVo data. We followed a hard-label approach, for each ExVo audio file we picked the categorical emotion based on the highest score provided by the raters to the corresponding emotion category. The network comprises of four convolutional layers followed by a fully connected layer with ten nodes and an output unit consisting of ten emotion classes. The output layer had softmax activation, while all the layers had ReLU activation. 
To train the neural network we split the train-set into training-subset and cross-validation subset in 90:10 ratio, this cross-validation subset was used for hyperparameter tuning. The ExVo provided validation set was used for inference. This training method made sure that the network trained for generating embeddings is trained in a speaker-independent fashion as per the baseline protocols~\cite{BairdExVo2022}. The networks were trained using cross-entropy loss with stochastic gradient descent. The learning rate was halved, in the range 10\textsuperscript{$-1$} to 10\textsuperscript{$-6$}, between successive epochs whenever the validation-loss stopped reducing. We used Keras deep learning library with tensorflow backend. \\ 
\underline{(5) Raw/Zff(ExVo)-CR}:
These systems are similar to the above mentioned system in all aspects apart from the fact that these have been optimised for the country-labels of ExVo data and were trained for a four-class classification task hence the output unit of the network consist of four country labels, the only architecture difference when compared to Raw/Zff(ExVo)-ER. \\
For both of these in-domain task-specific trained systems, the frame-level neural embeddings of dimension 10, are extracted before the activation layer of the fully-connected layer. A fixed-length utterance-level representation is obtained by computing functionals (mean and standard deviation) of the frame-level neural embeddings, denoted by Raw/Zff(ExVo)-ER and Raw/Zff(ExVo)-CR for (4) and (5) respectively. This makes these embeddings of dimension 20 (10 mean $+$ 10 std) each. For the purpose of analysis we provide the hard-classification based results for these four systems in Table~\ref{tab:classification_results} trained to classify emotions and country using the ExVo multi-task dataset. 
\vspace*{-0.25cm}
\subsection{Multi-task learner}
\vspace*{-0.2cm}
For a fair comparison and analysis of our derived embeddings to the challenge baseline features and results, we used the ExVo Multi-task baseline network \cite{BairdExVo2022} for that takes the fixed length representations as input and generates the predictions. The multi-task baseline model is a neural-network with two hidden layers, with 128 neurons in the first hidden layer followed by 64 neurons in the second layer, with leaky-ReLu activation used for both the layers. Adam optimization method \cite{kingma2014adam} is used for training the network, three loss-functions are used (1) the Mean Squared Error (MSE) for the age and emotion detection sub-task and (2) the cross-entropy loss for the native-country prediction sub-task. The mean of these loss functions is calculated for the target.  We developed systems with two different configurations: (a) `\textbf{Sys-1}' where the network configuration is same as the organizers, and (b) `\textbf{Sys-2}' where we doubled the number of neurons in the hidden layers that is 256 neurons for the first hidden and 128 neurons for the second.
 
Beside studying the embeddings individually, we also investigated (a) early fusion of the embeddings by concatenating the fixed-length representations and (b) hybrid system where predictions of different systems are fused to generate the final output for evaluation.

\begin{table}[!t]
    \centering
    \resizebox{\columnwidth}{!}{\begin{tabular}{ccccccc} 
{\bf Systems} & {\bf Dims.}  & {\bf Config.}& {\bf Emo-CCC}& {\bf Cou-UAR}& {\bf Age-MAE}& {\bf $S_{MLT}$} 
\\ \hline
\multicolumn{7}{c}{ \textsc{\bf ExVo Baseline on Validation-data} } \\ \hline 
\textsc{ComParE} & 6373 &  Sys.1 & 0.416 & 0.506 & 4.222 & 0.349 $\pm$ 0.003   \\
\textsc{Deep Spectrum}& 4096 & Sys.1 & 0.369 & 0.456 & 4.413 & 0.322 $\pm$ 0.003   \\ \hline
\multicolumn{7}{c}{ \textsc{\bf Self-Supervised (Pre-trained) on Validation-data} } \\ \hline 
\textsc{WavLM} & 2048  & Sys-1 & 0.523 & 0.542 & 4.094 & 0.382 $\pm$ 0.006  \\ 
\textsc{WavLM} & 2048  & Sys-2 & 0.548 & 0.536 & 4.008 & 0.390 $\pm$ 0.009  \\
\textsc{HuBERT} & 2048  & Sys-1 & 0.513 & 0.508 & 3.864 & 0.385 $\pm$ 0.004 \\
\textsc{HuBERT} & 2048  & Sys-2 & 0.518 & 0.508 & 3.782 & 0.391 $\pm$ 0.010 \\
\textsc{Wav2Vec2} & 1536  & Sys-1 &  0.390 & 0.379 & 3.903 & 0.330 $\pm$ 0.008 \\
\textsc{Wav2Vec2} & 1536  & Sys-2 &  0.385 & 0.376 & 3.895 & 0.328 $\pm$ 0.005 \\

\hline
\multicolumn{7}{c}{ \textsc{\bf Raw-Wav (Pre-trained for ASR and SER) on Validation-data} } \\ \hline 
\textsc{Raw (SER)} & 20  & Sys-1 & 0.078 & 0.342 & 3.909 & 0.153 $\pm$ 0.004 \\
\textsc{Raw (SER)} & 20  & Sys-2 & 0.084 & 0.332 & 4.032 & 0.158 $\pm$ 0.004 \\
\textsc{Zff (SER)} & 20  & Sys-1 & 0.083 & 0.320 & 4.021 & 0.156 $\pm$ 0.021  \\
\textsc{Zff (SER)} & 20  & Sys-2 & 0.087 & 0.335 & 3.897 & 0.163 $\pm$ 0.003  \\
\textsc{Raw (ASR)} & 2048  & Sys-1 & 0.370 & 0.477 & 4.210 & 0.333 $\pm$ 0.002 \\ 
\textsc{Raw (ASR)} & 2048  & Sys-2 & 0.392 & 0.465 & 4.179 & 0.338 $\pm$ 0.005 \\ 

\hline
\multicolumn{7}{c}{ \textsc{\bf Raw-Wav trained systems on Validation-data } } \\ \hline 
\textsc{Raw (ExVo)-ER} & 20 & Sys-1 & 0.454 & 0.331 & 3.953 & 0.327 $\pm$ 0.006  \\
\textsc{Raw (ExVo)-ER} & 20 & Sys-2 & 0.469 & 0.328 & 3.805 & 0.334 $\pm$ 0.005  \\
\textsc{Zff (ExVo)-ER} & 20 & Sys-1 & 0.385 & 0.330 & 3.886 & 0.315 $\pm$ 0.007 \\
\textsc{Zff (ExVo)-ER} & 20 & Sys-2 & 0.386 & 0.333 & 3.868 & 0.317 $\pm$ 0.006 \\

\hline
\multicolumn{7}{c}{ \textsc{\bf Early Fusion Experiments on Validation-data} } \\ \hline 
\textsc{Raw (ExVo)-ER +Zff (ExVo)-ER} & 40  & Sys-1 & 0.380 & 0.335 & 3.888 & 0.316 $\pm$ 0.009 \\
\textsc{Raw (ExVo)-ER + Zff (ExVo)-ER} & 40  & Sys-2 & 0.387 & 0.334 & 3.880 & 0.317 $\pm$ 0.008 \\
\textsc{Raw (ExVo)-ER +Raw (ASR)} & 2068  & Sys-1 & 0.440 & 0.480 & 4.159 & 0.352 $\pm$ 0.003 \\
\textsc{Raw (ExVo)-ER +Raw (ASR)} & 2068  & Sys-2 & 0.452 & 0.488 & 4.144 & 0.357 $\pm$ 0.003 \\
\textsc{Raw (ExVo)-ER + WavLM} & 2068 & Sys-1 & 0.546 & 0.542 & 4.150 & 0.383 $\pm$ 0.006 \\
\textsc{Raw (ExVo)-ER + WavLM} & 2068 & Sys-2 & 0.556 & 0.522 & 4.057 & 0.386 $\pm$ 0.004 \\
\hline
\multicolumn{7}{c}{ \textsc{\bf Hybrid Experiments on Validation-data} } \\ \hline 
\textsc{Raw (ExVo)-ER + Raw (ExVo)-CR} & 40 & Sys-1 & 0.454 & 0.437 & 3.953 & 0.355   \\
\textsc{Raw (ExVo)-ER + Raw (ExVo)-CR} & 40 & Sys-2 & 0.469 & 0.437 & 3.805 & 0.365   \\
\textsc{Raw (ExVo)-ER + Zff (ExVo)-CR} & 40 & Sys-1 & 0.454 & 0.456 & 3.953 & 0.359   \\
\textsc{Raw (ExVo)-ER + Zff (ExVo)-CR} & 40 & Sys-2 & 0.469 & 0.456 & 3.805 & 0.369   \\
\hline
\hline
\end{tabular}
}
\caption{Experimental results based on validation set. Abbreviations: ER - emotion recognition, CR - country recognition.\\}
\vspace*{-0.7cm}
\label{tab:table_validation_Results}
\end{table}

\begin{table}[!htb]
    \centering
    \resizebox{\columnwidth}{!}{\begin{tabular}{ccccccc} 
{\bf Systems} & {\bf Dims.} & {\bf Config.}& {\bf Emo-CCC}& {\bf Cou-UAR}& {\bf Age-MAE}& {\bf $S_{MLT}$} \\ \hline
\multicolumn{7}{c}{ \textsc{\bf ExVo Baseline on Test-data} } \\ \hline 
\textsc{ComParE}                & 6373 &  Sys-1 & 0.427 & 0.473 & 4.502 & 0.335   \\
\textsc{Deep Spectrum}          & 4096 & Sys-1 & NA & NA & NA & 0.305    \\
\hline
\multicolumn{7}{c}{ \textsc{\bf Proposed systems on Test-data} } \\ \hline
\textsc{WavLM}                  & 2048 & Sys-2 & 0.546 & 0.52 & 4.389 & 0.3684 \CheckmarkBold  \\
\textsc{Raw (ExVo)-ER}             & 20 & Sys-2 & 0.472 & 0.321 & 4.160 & 0.319  \\
\textsc{Raw (ASR)}              & 2048 & Sys-2 & 0.387 & 0.413 & 4.469 & 0.3167 \\ 
\textsc{Raw (ExVo)-ER + Raw (AMI)} &  2068 & Sys-2 & 0.461 & 0.412 & 4.493 & 0.3301 \\
\textsc{Hybrid}                 & 2068 &  Sys-2 & 0.546 & 0.52 & 4.160 & \textbf{0.379} \CheckmarkBold   \\
\hline
\hline
\end{tabular}
}
\caption{Experimental results of five selected systems on test-set. (\CheckmarkBold) indicate the systems outperforming the baseline results.}
\vspace*{-0.7cm}
\label{tab:table_test_Results}
\end{table}

\vspace*{-0.2cm}
\section{Results and Analysis}
\label{sec:Results}
\vspace*{-0.3cm}
Evaluation scores on validation-data are provided in Table~\ref{tab:table_validation_Results} whereas scores for the test-data based on five preferred systems are presented in Table~\ref{tab:table_test_Results}.

\vspace*{-0.2cm}
\subsection{Result analysis}
\vspace*{-0.2cm}
\underline{SSL embeddings}: It can be observed from Table~\ref{tab:table_validation_Results} that WavLM and HuBERT provide the best overall results ($S_{MLT}$), where WavLM performs better for Emotion and country sub-task whereas HuBert scores well on the age sub-task, Sys-2 configuration enhance the performance for both these cases. Wav2Vec2 optimises well for the age subtask but is inferior for others and Sys-2 configuration degrades its overall performance. \\
\underline{Out-of-domain task-specific embeddings}: In Table~\ref{tab:table_validation_Results}, the Raw/Zff(SER) shows overall inferior performance, but beats the baseline for the age sub-task, where Zff(SER) Sys-1 works particularly well on the age sub-task. Raw(ASR) performs equivalent to Wav2Vec2 for emotion sub-task and shows superior performance for the country sub-task. The overall score for Raw(ASR) are comparable to the baseline score where it beats the DeepSpectrum and even Wav2Vec2 from the SSL sub-group.\\
\underline{In-domain task-specific embeddings}: Raw(ExVo)-ER outperforms the baseline ComPARE features, for emotion and age sub-task in both validation and test set but does not perform well on the country sub-task. But from Table~\ref{tab:classification_results} it can be observed that Raw(ExVo)-CR and Zff(ExVo)-CR in particular outperforms the baseline method for the country sub-task on the validation-set.\\
\underline{Early-fusion based embeddings}: From the result of Raw(ExVo)-ER $+$ Raw(ASR) in Table~\ref{tab:table_validation_Results} it could be observed that these embeddings complement each other. When fused this system provides superior overall score than the baseline, and Sys-2 configuration further improves the performance while for other cases early-fusion doesn't offer much benefits.\\
\underline{Hybrid systems}: task specific predictions, emotion and age from Raw/Zff(Exvo)-ER and country predictions from Raw/Zff(ExVo)-CR yields an overall score that outperforms the validation-set baseline, and reduces the overall score-gap with the best performing SSL system.

As Sys-2 configuration systematically yielded improvements on the validation set, we evaluated the systems on the test set using that configuration. From Table~\ref{tab:table_test_Results}, it can be observed that the trends on the test set is similar to the trends observed on the validation set. The WavLM system yields superior performance for emotion and country sub-task while Raw(ExVo)-ER yields superior performance for the age task. Furthermore, in the hybrid approach, it can be observed that combination of the outputs of the two system yields the best system, which outperforms the ComPARE baseline score ($S_{MLT}$) with a relative 13\% margin.


\vspace*{-0.2cm}
\subsection{Filter analysis}
\begin{figure}[!htb]
	\centering
	\includegraphics[trim={0cm 4.7cm 0cm 0cm},clip=true,width=0.5\textwidth]{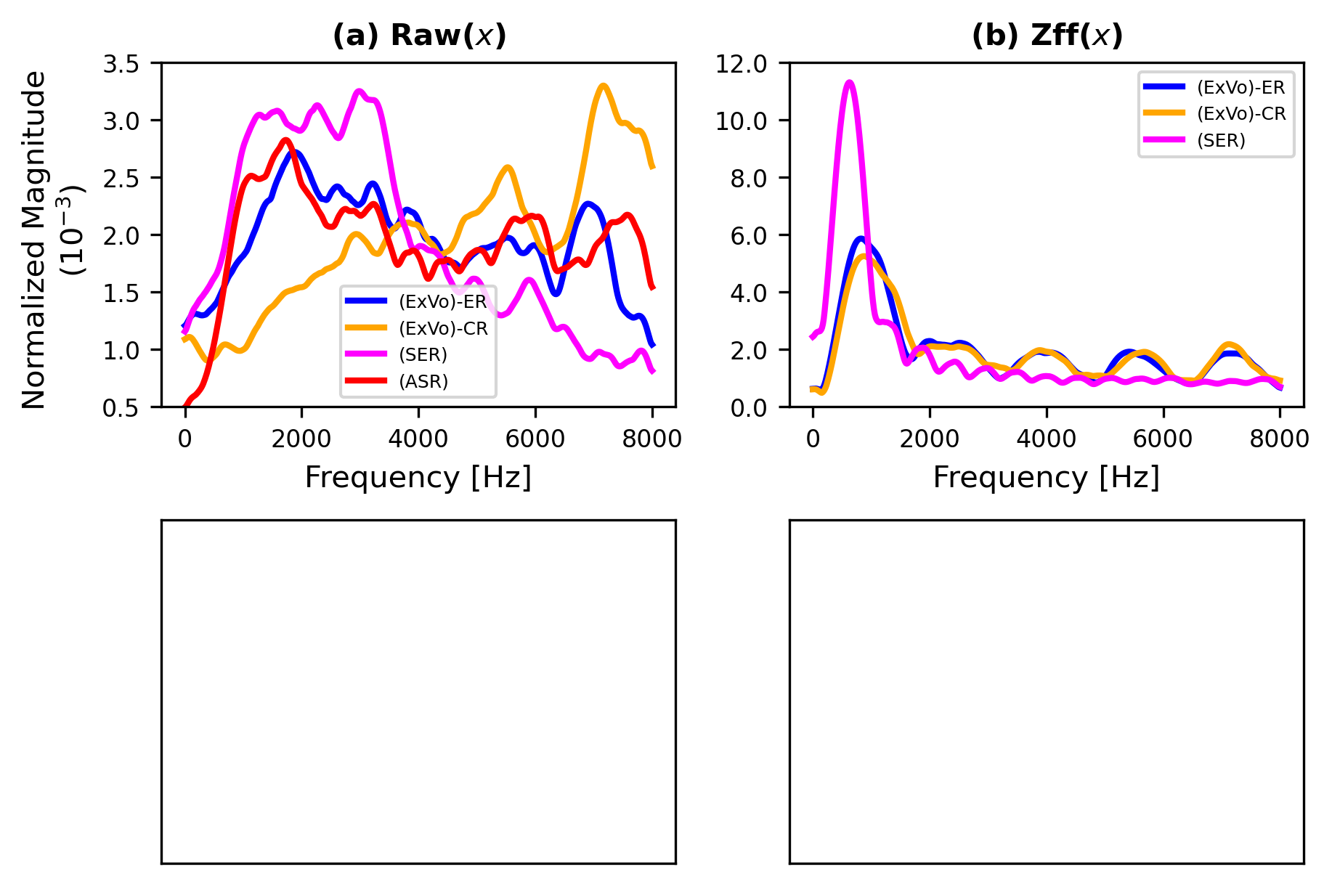}
	\vspace{-0.5cm}
	\caption{Cumulative frequency response of the first convolution layer for task-specific networks. (a) Raw ($x$) and (b) Zff ($x$) ; $x$ denotes the task.}
	\label{fig:subseg_response}
	\vspace{-0.7cm}
\end{figure}

\vspace*{-0.2cm}
The cumulative frequency response of the first convolutional layer for all the task-specific supervised networks are presented in Figure~\ref{fig:subseg_response}. From Figure~\ref{fig:subseg_response}(a), it is interesting to observe that the frequency response of out-of-domain pre-trained network - Raw(ASR), and in-domain trained network- Raw(ExVo)-ER, follows a similar trend that emphasizes on a wide frequency range and the $S_{MLT}$ score for these networks are comparable as well, whereas for Raw(SER) the filter emphasise is more on 1000-4000 Hz frequency region. For Raw(ExVo)-CR the trend shifts more towards high frequency. For Zff Figure~\ref{fig:subseg_response}(b) the trend is always towards the lower frequency specially for Zff(SER) whereas for Zff(ExVo)-ER/Cr there is some emphasis given to higher frequency ranges as well.
\vspace*{-0.5cm}

\vspace*{-0.4cm}
\section{Conclusion}
\label{sec:conclusion}
\vspace*{-0.2cm}
We investigated different neural embeddings for the ICML ExVo-2022 Multi-task learning challenge. Our investigations showed that SSL-based representations typically yield better overall score($S_{MLT}$) than feature representations obtained by training neural networks in a task-dependent manner. The improvements are largely observed in terms of Emo-CCC and Cou-UAR. When comparing out-of-domain task-dependent feature representation, we observe that representations obtained from neural network trained to classify phones yields better system than neural network trained for SER. Interestingly, we observed that in-domain task-dependent feature representation learning, i.e., Raw(ExVo)-ER tends to capture information similar to the phone classification task network and yields comparable system. Finally, our studies demonstrated that task-specific networks optimise well for the tasks. They can surpass the systems based on hand-crafted features and can complement SSL feature representations.



\textbf{\underline{Acknowledgements}:} This work was partially funded by the Swiss Science National Foundation through the Bridge Discovery project EMIL: Emotion in the loop - a step towards a comprehensive closed-loop deep brain stimulation in Parkinson's disease (grant no. $40B2-0{\_}194794/1$).


\nocite{langley00}

\bibliography{example_paper}
\bibliographystyle{icml2022}




\end{document}